\documentclass[fleqn,10pt]{wlscirep}
\usepackage[utf8]{inputenc}
\usepackage[T1]{fontenc}
\usepackage{soul}
\usepackage{subfiles}

\title{Cues to gender and racial identity reduce creativity in diverse social networks}

\author[1]{Raiyan Abdul Baten}
\author[2]{Richard Aslin}
\author[3]{Gourab Ghoshal}
\author[4,1,*]{Ehsan Hoque}
\affil[1]{Department of Electrical and Computer Engineering, University of Rochester, NY, USA}
\affil[2]{Haskins Laboratories and Department of Psychology, Yale University, CT, USA}
\affil[3]{Department of Physics and Astronomy, University of Rochester, NY, USA}
\affil[4]{Department of Computer Science, University of Rochester, NY, USA}

\affil[*]{mehoque@cs.rochester.edu}


\begin{abstract}
The characteristics of social partners have long been hypothesized as influential in guiding group interactions. Understanding how demographic cues impact networks of creative collaborators is critical for elevating creative performances therein. We conducted a randomized experiment to investigate how the knowledge of peers' gender and racial identities distorts people's connection patterns and the resulting creative outcomes in a dynamic social network. Consistent with prior work, we found that creative inspiration links are primarily formed with top idea-generators. However, when gender and racial identities are known, not only is there (1) an increase of $82.03\%$ in the odds of same-gender connections to persist (but not for same-race connections), but (2) the semantic similarity of idea-sets stimulated by these connections also increase significantly compared to demography-agnostic networks, negatively impacting the outcomes of divergent creativity. We found that ideas tend to be significantly more homogeneous within demographic groups than between, taking away diversity-bonuses from similarity-based links and partly explaining the results. These insights can inform intelligent interventions to enhance network-wide creative performances.
\end{abstract}
\begin{document}

\flushbottom
\maketitle

\thispagestyle{empty}

\section*{Introduction}
\begin{quote}{\it ``Outstanding discoveries, insights and developments do not happen in a vacuum.''}---Nobel Laureate Tom Steitz, on the impacts of discussions with collaborators on his research
\end{quote}
Creativity, although an individual strength,  benefits strongly from interactions with others. Online/offline networks of creative collaborators, e.g., among researchers, designers, or marketers, can socially stimulate novel ideas in people~\cite{azoulay2010superstar}. The network dynamics of who seeks creative inspiration from whom can therefore critically impact the creative outcomes of a population. In such social settings, people inevitably and implicitly respond to the demographic cues of their peers~\cite{poleacovschi2021gendered,zinn2019truly,richard2020effects}. Whether the knowledge of peers' demographic identities distorts the dynamics of the network ties and, consequently, the associated creative outcomes, remain important questions for both scientific and practical interests. While there exists a body of literature examining the diversity effects on creativity at individual and group levels~\cite{hofstra2020diversity,page2019diversity,nielsen2017opinion,shin2012cognitive,bassett2005paradox}, such studies typically overlook the temporally dynamic nature of real social networks, limiting our scientific understanding of its implications. Previous work shows that the dynamic nature of social networks can influence human cooperation~\cite{rand2011dynamic}, collective intelligence~\cite{almaatouq2020adaptive}, and the development of public speaking skills~\cite{shafipour2018buildup}, among other examples, but the effects in the creativity context remain unclear. From a practical perspective, it is desirable to elevate the creative capabilities of human networks. As automation takes away manual and repetitive jobs, the future of work will increasingly require offline/online ensembles of people to be adept in social-cognitive avenues of soft-skills such as creativity~\cite{frank2019toward,manyika2017future,baten2019upskilling}. Understanding the effects of demographic cues on networked creativity is critical for designing informed interventions that can promote creative outcomes in both offline (e.g., large-scale creative teams) and online (e.g., in work-from-home settings, social media) dynamic networks alike.  

There are two competing hypotheses on how social networks might adapt to demographic signals when people seek creative stimulation from peers. On the one hand, there is the notion of a {\it diversity bonus} that can help creative ideation~\cite{page2019diversity}. People from different demographic identities (e.g., gender/race) come with different concerns, perspectives, and life experiences. This can impact the knowledge domains they have access to, influencing the ideas they generate~\cite{hofstra2020diversity}. Furthermore, humans are known to be exceptionally adept at employing cues of skill/competence, success, prestige, age, health, and self-similarity to assess who most likely possess information useful to them, and form social links accordingly~\cite{henrich2015secret,henrich2015big,boyd2011cultural,herrmann2007humans}. Following this, it seems promising for (rational) humans to selectively form links with people from different demographic identities, so as to cash in on the diversity bonuses to generate novel ideas. 

On the other hand, the social network literature has extensively documented {\it homophily}, where people tend to form and sustain links with peers similar to themselves~\cite{mcpherson2001birds}. It can then be argued that people might forego potential diversity bonuses in favor of self-similarity-based network connections. Previous work suggests that if multiple people draw inspirations from similar stimuli, then even their independently stimulated ideas can become similar to each other~\cite{baten2020creativity}. This way, homophily-guided network dynamics can limit the variety of ideas stimulated---leading to counter-productive outcomes in divergent creativity~\cite{runco2014creativity}. An implicit conflict thus results from the trade-off between the diversity bonus and homophily---choosing to affiliate with others based on self-similarity attenuates creativity while the collective effect of group diversity enhances creativity.

Unfortunately, despite its scientific and practical importance, our understanding of the impact of demographic cues on networked creativity is limited, because such a test requires independently controlling for the exposure of the social-identity cues within an evolving social system. Furthermore, it is challenging to identify a dataset in the wild that allows for traceable links between the generation of creative ideas and the social context that facilitates them, gives agency to the participants for choosing their own inspiration sources, captures temporal evolution information of the dynamic network, and offers a way of revealing demographic information without introducing confounding effects. We therefore design and conduct a randomized controlled experiment within the virtual laboratory. Using a custom social-media-like web-interface for the interactions, we are able to unambiguously track creative stimulation of ideas in a dynamic social network, while the challenges of ensuring an unconfounded exposure to demographic cues are overcome with the use of avatars. 

\begin{figure*}
    \centering
    \includegraphics[width=0.95\linewidth]{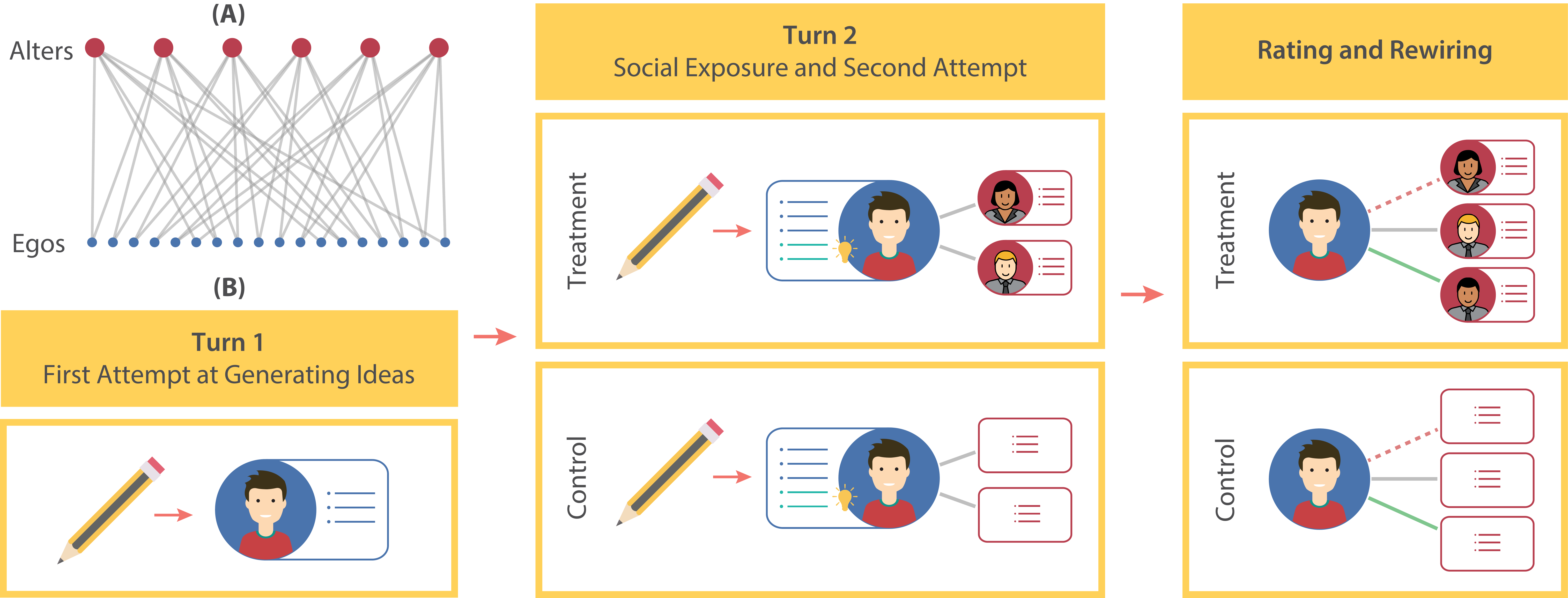}
    \caption{\textbf{Experimental Setup.} \textit{(A)} The bipartite network used as the initial configuration. Pre-recorded ideas of the alters were shown to the egos of both study conditions. Each ego was connected to $2$ alters. \textit{(B)} The study protocol for each of the $5$ rounds. In turn-$1$, the egos generated ideas independently. Turn-$2$ involved an additional idea-generation task on the same object. In this turn, the control egos were shown only the ideas of their alters for inspiration, while both the ideas and the demographic information of the alters were shown to the treatment egos. The egos could add inspired ideas to their lists. Finally, the egos rated the ideas of all the alters in the trial, and optionally updated the $2$ followee-alters at the end of each round. In the illustration, the ego dissolved the red dashed link to form the green link, while the grey link persisted.}
    \label{protocol_comb}
\end{figure*}

\section*{Experimental Setup}
Participants recruited from Amazon Mechanical Turk took part in $5$ rounds of text-based creative tasks, where they generated alternative use ideas for a given common object (e.g., a brick) in each round. We adopt a bipartite network design that involved two kinds of roles for the participants: as \textit{alters} ($N=12$) and as \textit{egos} ($N=180$). The alters' ideas were pre-recorded to be used as stimuli for the egos. The egos were randomly placed into either of two conditions: (1) Control ($N=90$, demographic cues not shown) and (2) Treatment ($N=90$, demographic cues shown). We ran two trials of the study. Each trial consisted of $6$ alters, whose ideas were shown to both the control and treatment egos in that trial.

Initially, each ego was randomly assigned to `follow' $2$ alters (out of $6$ in the trial). In each round, the egos first generated ideas independently (turn-1). In the control condition, the egos were then shown the ideas of the $2$ alters they were following. However, in the treatment condition, the egos were additionally shown the demographic information (gender and race) of their followee alters using avatars. If the egos got inspired with new alternate use ideas on the same object, they could add those ideas to their own lists (turn-2). Then, the egos were shown the ideas of all $6$ alters, which they rated on novelty. Finally, they were allowed to optionally follow/unfollow alters to have an updated list of $2$ followee alters each. In turn-2 of the following round, they were shown the ideas of their newly chosen alters (Figure~\ref{protocol_comb}; Methods). Importantly, the only difference between the two conditions was that the treatment egos had access to the gender and race cues of the alters. Thus, any difference in the connectivity dynamics and network-level creative outcomes between the two conditions can be attributed to the availability of demographic cues.

\section*{Results}
\subsection*{Same-gender links are highly stable in the presence of demographic cues} 
\begin{figure}
    \centering
    \includegraphics[width=0.45\linewidth]{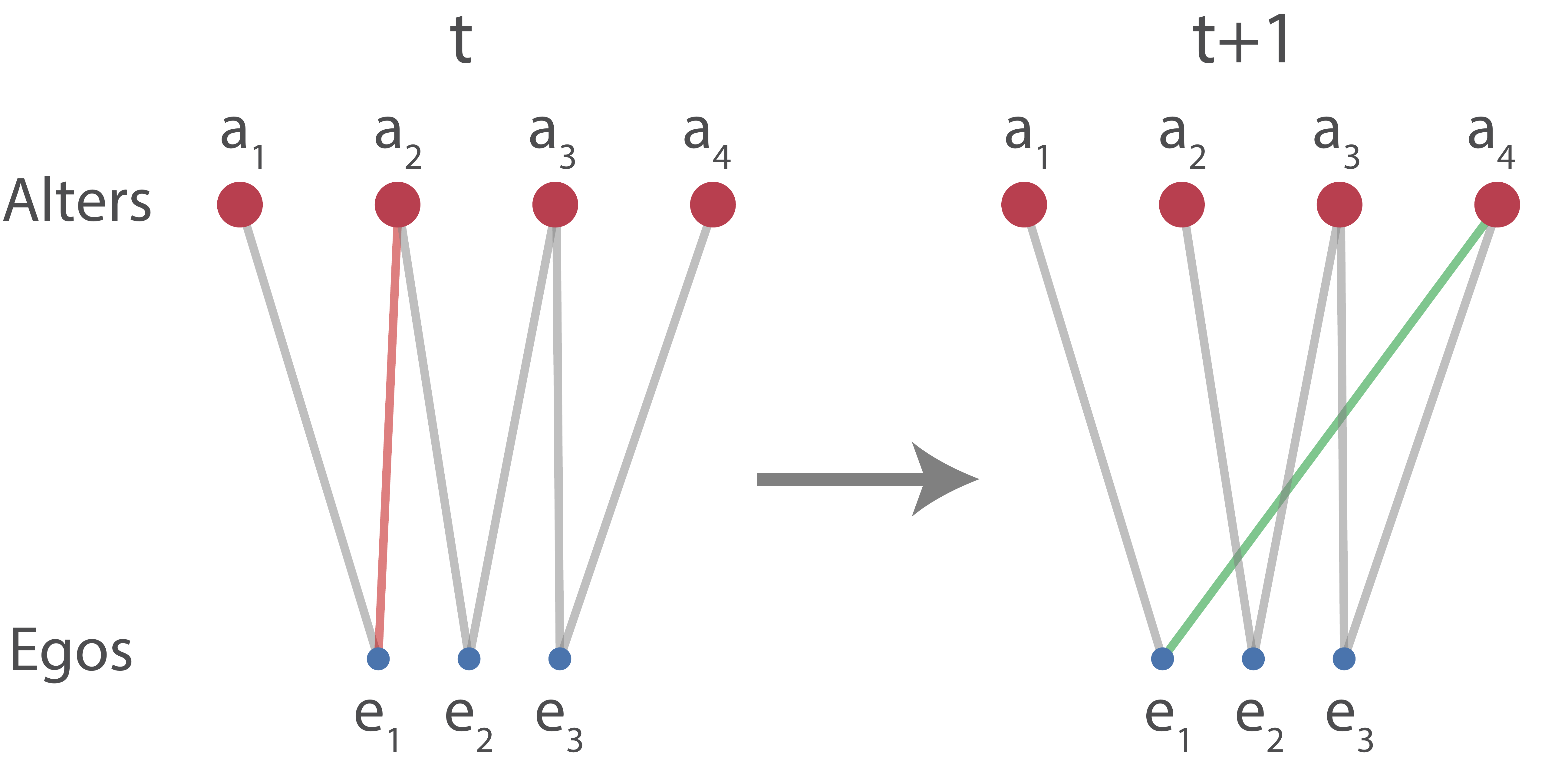}
    \caption{\textbf{Intuition behind STERGMs.} To capture the formation and persistence dynamics of the links, two separate models are fitted for the temporal networks of each study condition. The formation model tracks links that do not exist in time $t$, but exist in time $t+1$, e.g., the green link between ego $e_1$ and alter $a_4$. The persistence model considers links that exist in both time instances, e.g., all grey links. Since the egos had to follow a constant number of two alters, the red link $e_1$-$a_2$ needed to dissolve to allow the green link to form. However, if a link does not persist, it must dissolve---thus, the dissolution effects are captured in the persistence model and we need not fit a separate model for dissolution.}
    \label{stergm}
\end{figure}

We employ Separable Temporal Exponential Random Graph Models~\cite{krivitsky2014separable} to capture the link dynamics in the temporal network data. Two separate models capturing link (1) Formation and (2) Persistence patterns are fitted for each study condition (Figure~\ref{stergm}). As exogenous features, we choose attributes that the treatment egos were likely to consider in making connectivity decisions: (a) round-wise creative performances of the alters (measured by non-redundant idea counts), (b) gender-based homophily, and (c) race-based homophily. Additionally, we employ one endogenous feature of edge-counts, to control for network density (see Methods for details).

In the control condition, we find that both the link formation and persistence patterns are significantly guided by the creative performances of the alters (formation model: $\beta = 0.324$, $Z$-value = $5.51$, $P<10^{-4}$; persistence model: $\beta = 0.417$, $Z$-value = $6.95$, $P<10^{-4}$). The positive $\beta$ values suggest that better performing alters are more likely to be followed by the egos and that these links are significantly stable across rounds. The gender and race features do not show any significant effect once the performance-based link dynamics are accounted for ($P>0.05$ for both features in both models). This is intuitive, as the egos did not have any information about the alters' gender and race, and could only see their ideas.

In the treatment condition, the link formation model once again shows only the creative performances of the alters to be a significant predictor ($\beta = 0.197$, $Z$-value = $3.93$, $P<10^{-4}$), and not the demographic features ($P>0.05$ for both). However, we observe a notably different trend in the link persistence model. We find that link persistence depends significantly on both non-redundant idea counts ($\beta = 0.355$, $Z$-value = $6.08$, $P<10^{-4}$) and gender-based homophily ($\beta = 0.599$, $Z$-value = $3.74$, $P<10^{-3}$). In other words, if a link exists between participants of the same gender, its odds of persisting increases by $82.03\%$, after controlling for merit-based persistence. No significant effect is observed for the race feature ($P>0.05$). 

In summary, the availability of demographic cues in the treatment condition is observed to be associated with a significant stability in same-gender links, unlike what is seen in the control condition (see SI Tables S1-S4). When presented with multiple modes of demographic information at the same time (e.g., both gender and race), humans in many contexts make choices based on the mode they find more salient~\cite{crisp2007multiple2}. This intuition can partly explain why we observe gender-based but not race-based link stability in our experiment.

\subsection*{Inter-ego semantic similarity increases when the alters' demographic cues are known} 
To assess the creative outcomes in the networks, we use natural language processing tools for computationally capturing the semantic qualities of ideas. Divergent thinking/creativity leads individuals to generate \textit{numerous} and \textit{varied} responses to a given prompt~\cite{runco2014creativity}. Two crucial dimensions of creativity, flexibility and originality, explicitly take into consideration the semantic qualities of ideas: flexibility is the
number of distinct semantic categories that a person accesses in his/her ideas, while originality is the extent to which a solution is semantically novel~\cite{rietzschel2007personal}. If the network processes systematically make the stimulated ideas of the participants semantically similar, it hurts the divergence purposes. We estimate the semantic nature of the egos' idea-sets using neural word embeddings~\cite{mikolov2013distributed}. To compute the semantic similarities between idea-sets, we employ the cosine similarity metric~\cite{jurafsky2000speech}. 

Previous work suggests that following the same people, i.e., having the same stimuli, can introduce semantic similarities among independently stimulated idea-sets~\cite{baten2020creativity}. Therefore, from each round, we collect pairs of egos who share (a) $2$ common alters (i.e., exactly the same stimuli), (b) $1$ common alter and (c) no common alter. Within these subgroups, we compute the semantic similarities between every ego-pair's stimulated ideas in turn-2 (see Methods).

\begin{figure}
    \centering
    \includegraphics[width=0.65\linewidth]{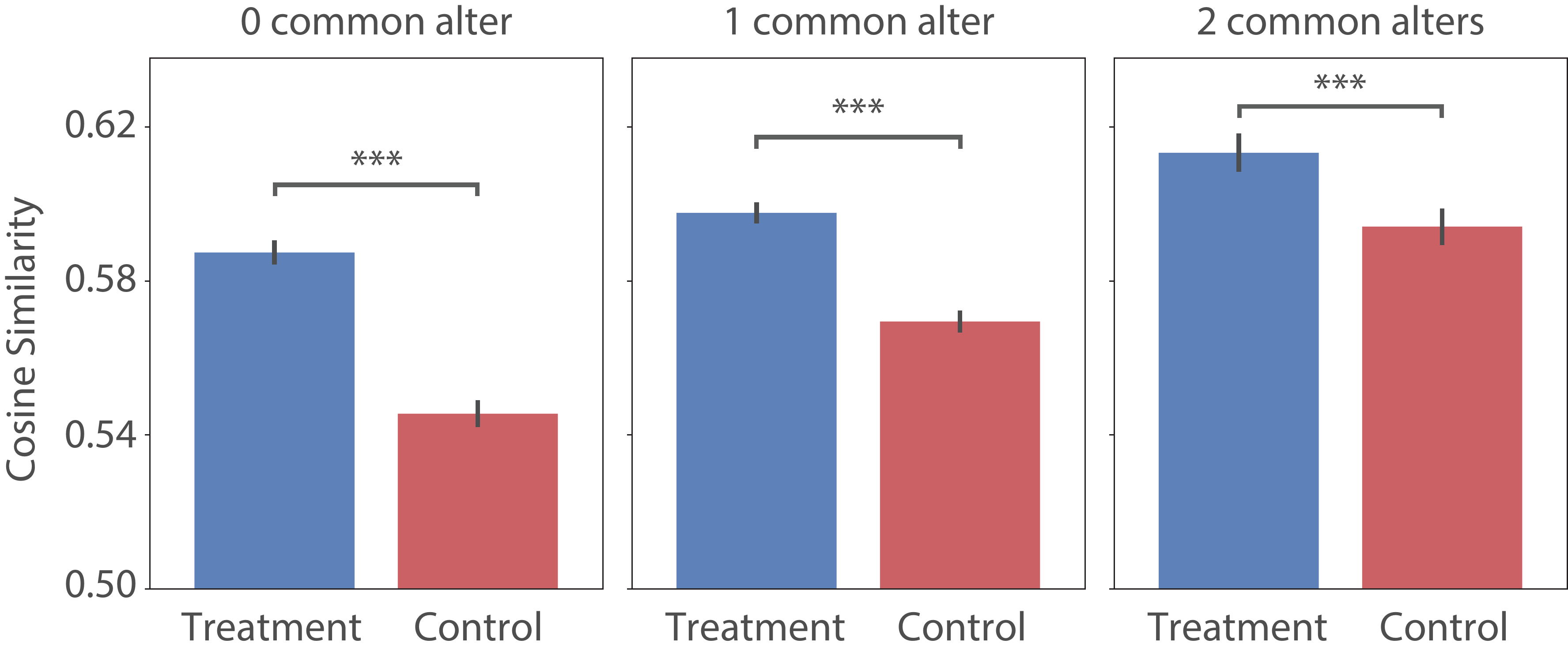}
    \caption{\textbf{Inter-ego semantic similarities under various conditions.} Cosine similarities between the idea-sets of egos-pairs are shown across three sub-groups: ego-pairs who share $0$, $1$ and $2$ common alters between them. With the increase in the number of common alters (i.e., increase in stimuli similarity), the inter-ego semantic similarity typically increases. Importantly, the semantic similarities are significantly higher in the treatment condition compared to the control condition in all three sub-groups. Whiskers denote $95\%$ C.I. ***$P<0.0001$, corrected for multiple comparisons.}
    \label{interego}
\end{figure}

Using a $3\times2$ factorial design ($3$ levels in the number of common alters and $2$ study conditions), we find significant main effects for both factors (Aligned Rank Transform procedure~\cite{wobbrock2011aligned}, $P<10^{-15}$ for both), and a significant interaction between them ($P<10^{-7}$; Figure~\ref{interego}). Post-hoc analyses reveal that in both the treatment and control conditions, the inter-ego similarities increase significantly as the number of common alters increase from $0$ to $1$ and also from $1$ to $2$ ($P<10^{-4}$ for all of these comparison cases).  These trends intuitively follow that inter-follower similarities can stem from having common stimuli. 

Notably, we observe that the inter-ego semantic similarities are significantly higher in the treatment condition compared to their control counterparts, in all of the common-alter-based subgroups ($P<10^{-4}$ in each). In addition, the inter-ego semantic similarities are found to increase with time (treatment: Pearson's $r=0.13$, $P<10^{-4}$; control: $r=0.12$, $P<10^{-4}$). These results mark systematically negative impacts of demographic similarity on the treatment networks' creative outcomes. Supplementary Text and Tables S5-S10 provide full details of the statistical analyses.

\subsection*{Homophily can make one's stimuli set less diverse}
\begin{figure*}
    \centering
    \includegraphics[width=1\linewidth]{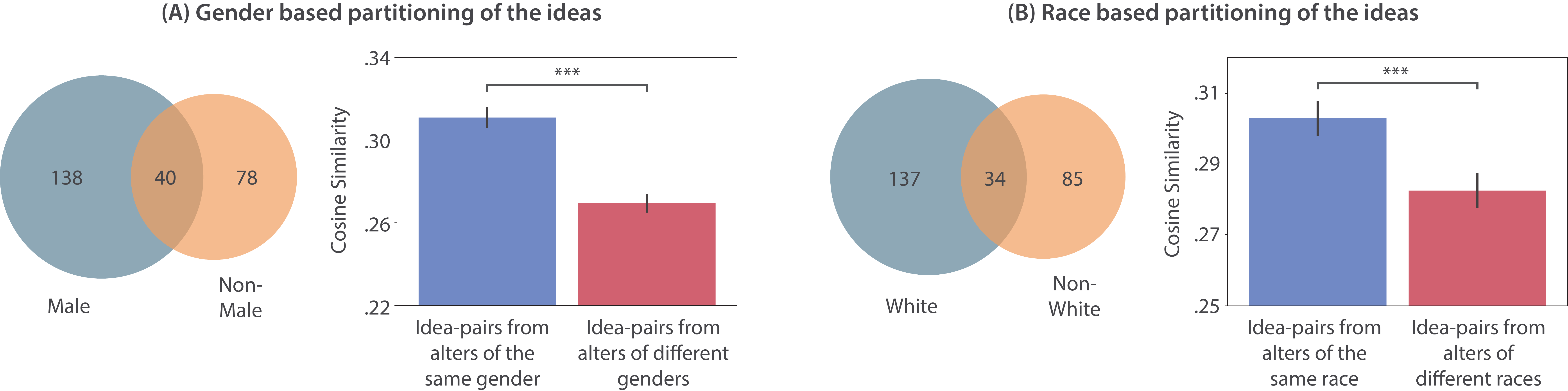}
    \caption{\textbf{Homogeneity of ideas within demographic-groups.} \textit{(A)} shows gender-based partitioning of the alters' ideas. Some of the ideas were submitted by both males and non-male alters, while others were submitted uniquely by either gender categories. We only consider ideas from the latter case in the similarity analysis. Pairwise comparisons of ideas within gender show a significantly higher similarity than idea-pairs between genders. \textit{(B)} shows the same insights in case of race-based partitioning of the alters' ideas. Whiskers denote $95\%$ C.I. ***$P<0.0001$.}
    \label{venn}
\end{figure*}

To explain these results, we test whether the well established intuition of ideas being more homogeneous within demographic groups than between~\cite{page2019diversity,burt2004structural} holds true for the alters' ideas. We find that idea-pairs within gender/race are indeed significantly more similar to each other than idea-pairs between gender/race ($2$-tailed tests; gender: $t(4633)=11.66$, $P<10^{-30}$; race: $t(4870)=5.73$, $P<10^{-7}$; Figure~\ref{venn}; Methods). Thus, it follows logically that homophily-guided network dynamics can make a follower's stimuli idea-set uniform and similar. This can deprive the follower of possible diversity bonuses, partly explaining the increased inter-ego semantic similarity observed in the treatment condition that can stem from having similar stimuli.

\section*{Discussion}
Our study design afforded several benefits over observational data. Crucially, in this design, the egos could dynamically update connections to alters every round, the links between ideas and their inspiration sources were unambiguously traceable, and the revelation of gender and race cues was controlled using avatars. This dynamic network approach is at a stark contrast to traditional diversity/group dynamics exploration setups, where the groups are typically treated as black boxes and the intra-group ties are seldom tracked explicitly~\cite{sawyer2011explaining}. Furthermore, our choice of a bipartite network structure ensured uniform stimuli-sets for the egos. However, like in all experiments, there were limitations, too. The unidirectional nature of the networks prohibited us from assessing the effects of natural bidirectional interaction settings. The avatars we used could potentially have differential effectiveness in indicating demographic contrasts between the gender and race of the alters, and the effectiveness could also vary for different egos. We nevertheless chose to use avatars over more realistic photos to standardize visual depictions and to remove confounding factors stemming from facial, personality, or other visual cues. We stopped collecting data for trial-2 when results consistent with trial-1 were achieved, thereby supporting our initial findings. However, future research should use a pre-specified sample size for trial-2 to provide a stronger replication of the trial-1 effects.

Diversity, creativity and homophily are complex bodies of knowledge, each driven by multidimensional mechanisms. For instance, a large body of work has examined how exposure to gender and race cues can bias people's behavior in various settings. Descriptions of identical medical symptoms from men and women can be perceived differently by physicians~\cite{colameco1983sex}, even leading to cases where men patients are more comprehensively investigated than women patients with the same symptoms~\cite{armitage1979response}. The perception of one's creativity can be biased based on identity as well: in their highly creative craft, male florists can be perceived as `naturally' and `truly' creative and thus held to higher regards than their female counterparts~\cite{zinn2019truly}. The power relations among various demographic identities have widely been documented and are known to dictate such biases. For example, the production of knowledge in academia can be racialized~\cite{thapar2019epistemic}, where the contributions by minorities can be systematically neglected~\cite{thapar2019epistemic,hofstra2020diversity}. Such identity-based realities can influence the occurrence of homophilic ties in many domains, for example in engineering, where knowledge can be perceived to be highly accessible when women seeks it from other women~\cite{poleacovschi2021gendered}.

Naturally, our work did not encompass every possible combination of contexts and scenarios that can emerge in such highly complex systems. Rather, we showed one set of empirical evidences in support of our arguments linking the interdisciplinary components, towards filling an important void in literature with regards to identity-informed creative inspiration seeking behavior in a temporal social network setting. We found evidence that demographic cues can indeed bias the connectivity dynamics in creativity-centric social networks, and systematically influence the creative outcomes therein. The more genetically diverse a bee hive, the higher is its survival probability~\cite{mattila2007genetic}. Likewise, a diversified pool of creative influencers can help people generate non-redundant ideas by allowing them to draw on diverse stimuli sets, something that is likely stifled in homophily-driven and highly centralized networks. 

Both of our study conditions had explicit instructions and incentives to generate creative ideas (elaborated in Methods), while only the treatment condition received implicit demographic cues. Previous literature suggests that such implicit priming can activate demography-specific schemas in people's memory~\cite{white2007race}, potentially transferring to how one approaches a creative task~\cite{gaither2015thinking}. Surface-level cues on gender or race can trigger the Social Categorization process~\cite{dahlin2005team}, whereby people differentiate themselves on the basis of demographic characteristics, e.g., race-based categorization can make the differences between White and Black more prominent. Our study revealed both gender and race cues simultaneously, which can trigger the Multiple Social Categorization process~\cite{crisp2007multiple2}. Under this process, shared identities can cut across the dichotomies of group identities. For example, if members belonging to Black and White categories are also female, then Black women and White women can be perceived as more similar compared to men~\cite{crisp2011cognitive}, reducing the race-based `us' versus `them' distinction. This process can partly explain why we observed a significant link persistence pattern in gender but not in race. Moreover, in the social science literature, both gender and race are known to be interwoven with questions of power and context. These identities often come with certain assumptions and stereotypes about social roles, including assumptions and stereotypes about knowledge: regarding who can know what and how the knowledge of different people are to be accessed, received, evaluated and prioritized~\cite{doan2017epistemic,dotson2014conceptualizing,settles2019scrutinized}. Such assumptions and stereotypes could also have played a role in guiding the egos' rewiring decisions, as the same-gender connections became significantly stable across time.

The Social Comparison process posits that people self-identifying to belong to different social categories can compete over identity and resources~\cite{sherif1961intergroup}. Given this, a potential sense of conflict can arise if the egos find their alter-sets to comprise entirely of `out-group' members, and feel compelled to choose creative stimulations from them. We guarded against this possibility by ensuring that we had at least two alters from every demographic dimension in each trial.

In the teamwork literature, such an `us-them' distinction is known to potentially reduce information-sharing among diverse members~\cite{horwitz2007effects}, often leading to fractures within the team~\cite{li2005factional}. Conversely, there can be benefits of the process, since the subgroups may enjoy reduced conflict, greater psychological safety, ease of communication~\cite{lau2005interactions}, and greater accessibility to knowledge~\cite{poleacovschi2021gendered}. The egos in our study merely viewed the stimuli and formed no proper `team' with the alters. Moreover, the ideas of the alters of various identities were equally accessible for the egos. This setting is rather similar to social media and academic networks of researchers, for example, where unidirectional and unreciprocated inspiration ties are common. Given this setting, the self-similarity based benefits, e.g., reduced conflict and psychological safety, were not on offer for the egos. We still saw a tendency of the egos to maintain links based on self-similarity of gender, thereby foregoing some values in diversity. It is not uncommon for people to report unreciprocated ties with creative others~\cite{mckay2017connected}; and we find that such ties extend to homophily-guided dynamics as well. In behavioral economics, it is well known that humans are not always perfectly logical, rational and informed actors, and can often diverge from optimal `rational' behavior (i.e., behavior that maximizes personal utility) in various social-psychological contexts~\cite{kahneman2011thinking}. Although maintaining ties with alters of different identities than one's own would seem to be the logical approach for the egos in our exploration, we observed the opposite trend of behavior. 

In summary, our results provide insights on how demography-driven behavior can influence creativity, a soft-skill with accelerated demand in an automation-driven world. We found that in a creativity-centric social network, same-gender links persisted significantly. Such behavior was not observed if the gender and race cues were not available to the participants, everything else held constant. Moreover, we found people's ideas to be more homogeneous within demographic groups than between---reinforcing the intuition that diversity bonuses might get compromised if one systematically maintains connections based on demographic identity. We indeed found that in the presence of demographic cues, the inter-ego semantic similarity increased significantly compared to the control condition where no such cue was shown, thus hurting divergent creativity.

These insights can help inform interventions/scaffolding in social systems where superior creative outcomes are sought after. For instance, due to the COVID-19 pandemic, many teams are brainstorming using remote-collaboration tools. While many face-to-face interaction benefits may go missing in such platforms, there can be upsides on offer, too. Following our findings, masking/modifying people's identities online can potentially help to elevate network-level creative outcomes. In social media, people often follow highly creative peers for novel ideas. Their choices of who to follow can be biased by demographic considerations. Algorithmic interventions can then be made to help people diversify their creative stimulation sources. Such measures can proactively guard against the inter-follower semantic similarities to enhance network-wide creative performances. In our future work, we intend to explore more nuanced research questions about the differential effects of identity attributes in epistemic burden, impacts and privileges in creativity-centric social interactions. In addition, we intend to extend the scope of the study from the binary classification of gender and race to a multi-class perspective, using higher-resolution data from a larger set of gender and race categories.

\section*{Methods}
\subsection*{Ethics Statement}
The study was reviewed and approved by the Institutional Review Board of the University of Rochester, NY, USA. All participants provided informed consent. All methods were carried out in accordance with relevant guidelines and regulations for involvement of human participants. 

\subsection*{Participants}
There were two trials in the study. In the first trial, the ideas of $6$ alters were used as stimuli for $72$ egos each in the control and treatment conditions. In the second trial, $6$ different alters acted as the stimulation sources for $18$ egos each in the two conditions. The second trial helped to ensure that our results do not overfit to the alters of the first trial, and gave results consistent with the first trial. Our reported results are derived from the full dataset from both of the trials together, to help reduce noise and increase the statistical power of the results. All of the participants were located in the United States. The participants chose for themselves pseudo usernames as per their liking, and were assigned their alter/ego roles and treatment/control conditions randomly.

Among the $180$ egos, $95$, $84$ and $1$ ego(s) respectively self-identified to be of male, female and other gender. $134$ of the egos self-identified as White, and $46$ of them as non-White. The age distribution was as follows: 18-24: 16, 25-34: 69, 35-44: 48, 45-54: 30, 55+: 17. As for the alters, in trial 1, $4$ of them identified as male and $2$ as female; while there were $3$ alters from each of the White and non-White race categories. In trial 2, there were $3$ alters from each of male and female genders; while $4$ of them identified as White and $2$ as non-White. The age distribution was: 18-24: 2, 25-34: 8, 35-44: 2. There is precedence in scientific literature of using binary splitting of data along gender (male vs. non-male) and race (White vs. non-White) dimensions, especially when the data availability is limited for many of the actual non-majority identity subgroups~\cite{jimenez2019underrepresented,hofstra2020diversity}. We adopted this binary-splitting strategy to make the statistical analysis feasible.

\subsection*{Measure of Divergent Creativity}
In this experiment, we are interested in divergent creativity, which deals with a person's ability to come up with or explore many possible solutions to a given problem~\cite{kozbelt2010theories}. We use a customized version of Guilford's Alternate Uses Test~\cite{guildford1978alternate}, the canonical approach for quantifying divergent creative performances\footnote{\normalfont \small Guilford's Alternate Uses Test is Copyright @ 1960 by Sheridan Supply Co., all rights reserved in all media, and is published by Mind Garden, Inc, www.mindgarden.com.}. In each of the $5$ rounds, the participants were instructed to consider an everyday object (e.g., a brick), whose common use was stated (e.g., a brick is used for building). The participants were asked to come up with novel and useful alternative uses for the object: uses that are different from each other, different than the given common use, and are appropriate and feasible. We choose the first $5$ objects from the Form B of Guilford's test as the prompt objects in the $5$ rounds.

\subsection*{Procedure} Each of turn-1 and turn-2 allowed the egos $3$ minutes to generate their ideas on the same object. In turn-2, the egos in the control condition were shown only the pseudo usernames and the lists of ideas of their followee alters. The egos in the treatment condition were additionally shown the gender (male and female) and race (White and non-White) information of the alters using text and avatars (see SI Figure S1). The avatars were used to ensure uniform visual depiction for all of the alters of the same demographic group, so as not to bias the egos by any facial, personality or other visual cues. The egos were instructed not to resubmit any of the alters' exact ideas, and told that only non-redundant ideas would contribute to their performance. They were also told that there will be a short test at the end of the study, where they will need to recall the ideas shown to them. This was to ensure that the participants paid attention to the stimuli ideas, which has been shown to positively impact ideation performances~\cite{nijstad2006group,dugosh2005cognitive,paulus2000groups,brown1998modeling}. After turn-2, the egos rated all the ideas of the $6$ alters in their trial on a 5-point Likert scale (1: not novel, 5: highly novel)~\cite{coursey2018divergent,bechtoldt2010motivated}. As the egos optionally rewired their network connections to have an updated list of which alters to follow, they were required to submit the rationale behind their choices of updating/not updating links in each round. This was in place to make the egos accountable for their choices, which has been shown to raise epistemic motivation and improve systematic information processing~\cite{bechtoldt2010motivated,scholten2007motivated}. The participants were paid \$10 upon the completion of the tasks, as well a bonus of \$5 if they were among the top 5 performers in groups of 18. SI Figures S2-S4 show the user interfaces.

\subsection*{Analysis Strategy}

\subsubsection*{Quantifying Creativity} Against the pool of ideas submitted by one’s peers, the number of non-redundant ideas that a participant comes up with is a widely accepted marker of his/her creativity~\cite{oppezzo2014give,abdullah2016shining}. The intuition being, to be creative, an idea has to be statistically rare. First, we filtered out inappropriate submissions that did not meet the requirements of being feasible and different from the given use. Then, all the ideas submitted in a given round by all the participants were organized so that the same ideas are binned or collected together. We followed the coding rules described by Bouchard and Hare~\cite{bouchard1970size} and the rules specified in the scoring key of Guilford's Alternate Uses test, Form B, for binning the ideas. On average, the egos submitted $5.43$ and $4.33$ ideas respectively in turn-1 and turn-2 of every round. These counts did not change significantly over time.

Once all the ideas were binned, we computed the non-redundant idea counts by looking at the statistical rarity of the ideas submitted by the participants. Namely, an idea was determined to be non-redundant if it was given by at most a threshold number of participants in a given pool of ideas. For the alters, the threshold was set to $1$, and the pools were set to be the round-wise idea-sets of the $6$ alters in the given trial.

The first author and two undergraduate research assistants independently coded the data to bin similar ideas together. The first author coded all of the ideas in the dataset, while the two research assistants binned ideas from a random split of $50\%$ participants each. The coders were shown the anonymized ideas in a random order. Based on their coding, the total non-redundant idea counts of the participants in all $5$ rounds were computed separately and the agreements were calculated. Between the first and second coder, the intra-class correlation coefficient was $ICC(3,2)=0.92$, $P<10^{-19}$, $95\%$ C.I. = $[0.88,0.95]$, and the Pearson's correlation coeffient was $r=0.87$, $P<10^{-20}$, $95\%$ C.I.=$[0.79,0.92]$. Between the first and third coder, the agreements were again high: intra-class correlation coefficient $ICC(3,2) = 0.88$, $P<10^{-15}$, $95\%$ C.I. = $[0.83,0.92]$; Pearson's $r=0.83$, $P<10^{-19}$, $95\%$ C.I.$=[0.74,0.89]$). The coded data from the first coder were then used in the analyses.

\subsubsection*{Capturing Link Formation and Persistence Dynamics: Separable Temporal ERGM} In the classic framework of the Exponential Random Graph Model (ERGM), the observed network (i.e., the data collected by the researcher) is regarded as one realization out of a set of possible networks originating from an unknown stochastic process we wish to understand. The range of possible networks, and their probability of occurrence under the model, is represented by a probability distribution on the set of all possible graphs with the same number of nodes as the observed network. Against these possible networks, we can then ask whether the observed network shows strong tendencies for structural characteristics that cannot be explained by random chance alone~\cite{robins2007introduction}. 

The basic expression for the classic (static) ERGM model can be written as,

\begin{equation}\label{ergm_vanilla}
    P(\mathbf{Y}=\mathbf{y} | \mathbf{X} = \mathbf{x}) = \Big( \frac{1}{\kappa}\Big) \text{exp} \{\mathbf{\beta}^T \mathbf{g}(\mathbf{y,x})\}
\end{equation}

Here, $\mathbf{Y}$ is the random variable for the state of the network (adjacency matrix), with a particular realization $\mathbf{y}$. $\mathbf{X}$ denotes the vector of exogenous attribute variables, while $\mathbf{x}$ is the vector of observed attributes. $\mathbf{\beta} \in \mathbb{R}^p $ is a $p \times 1$ vector of parameters. $\mathbf{g}(\mathbf{y},\mathbf{x})$ is a $p$-dimensional vector of model statistics for the corresponding network $\mathbf{y}$ and attribute vector $\mathbf{x}$. $\kappa$ is a normalizing quantity which ensures that Eq.~\ref{ergm_vanilla} is a proper probability distribution. Unfortunately, evaluating $\kappa$ exactly is non-trivial. Therefore, we need to resort to numerical methods to approximate the coefficients $\hat{\mathbf{\beta}}$. Namely, we use Markov Chain Monte Carlo methods to simulate draws of $\mathbf{Y}$, and from those draws we estimate the coefficients using Maximum Likelihood Estimation (MCMC-MLE method). Such estimation methods make it convenient to transform Eq.~\ref{ergm_vanilla} to the following equivalent conditional log-odds form:

\begin{equation}\label{ergm_conditional}
    \text{log}\Bigg[ \frac{P(Y_{ij}=1|\mathbf{y}^C_{ij},\mathbf{X})}{P(Y_{ij}=0|\mathbf{y}^C_{ij},\mathbf{X})}\Bigg] = \mathbf{\beta}^T \Delta_{ij}(\mathbf{y,x})
\end{equation}

Here, $\mathbf{y}^C_{ij}$ denotes all the observations of ties in $\mathbf{y}$ except $y_{ij}$. $\Delta_{ij}(\mathbf{y,x})$ is the \textit{change statistic}, which denotes the change in the value of the network statistic $\mathbf{g}(\mathbf{y,x})$ when $y_{ij}$ changes from $1$ to $0$. This emphasizes the log-odds of an individual tie conditional on all other ties.

For our temporal network data, we employed an extension of the static ERGM that deals with dynamic networks in discrete time: the Separable Temporal ERGM (STERGM). In contrast to static ERGMs, here we fitted two models: one for the underlying relational \textit{formation}, and another for the relational \textit{persistence}. In going from a network $\mathbf{Y}^t$ at time $t$ to a network $\mathbf{Y}^{t+1}$ at time $t+1$, the formation and persistence of ties are assumed to occur independently of each other within each time step (hence `separable'), to be captured by the two models respectively. The governing equations for the formation and persistence models, analogous to Eq.~\ref{ergm_conditional}, are then written respectively as:

\begin{equation}\label{formation}
    \text{log}\Bigg[ \frac{P(Y_{ij,t+1}=1|\mathbf{y}^C_{ij},\mathbf{X},Y_{ij,t}=0)}{P(Y_{ij,t+1}=0|\mathbf{y}^C_{ij},\mathbf{X},Y_{ij,t}=0)}\Bigg] = \beta_{f}^T \Delta_{ij,f}(\mathbf{y,x})
\end{equation}

\begin{equation}\label{persistence}
    \text{log}\Bigg[ \frac{P(Y_{ij,t+1}=1|\mathbf{y}^C_{ij},\mathbf{X},Y_{ij,t}=1)}{P(Y_{ij,t+1}=0|\mathbf{y}^C_{ij},\mathbf{X},Y_{ij,t}=1)}\Bigg] = \beta_{p}^T \Delta_{ij,p}(\mathbf{y,x})
\end{equation}

Here, time indices have been added to the equations unlike before, as well as new conditionals. In the formation model in Eq.~\ref{formation}, the expression is conditional on the tie not existing at the previous time step, whereas in the persistence model in Eq.~\ref{persistence}, it is conditional on the tie existing. Figure~\ref{stergm} summarizes these intuitions. There are separate coefficient vectors $\beta_f$ and $\beta_p$ for the formation and persistence models respectively, as well as separate change statistics $\Delta_{ij,f}(\mathbf{y,x})$ and $\Delta_{ij,p}(\mathbf{y,x})$ for the two models. Note that in the literature, it is common to refer to the persistence model as the `dissolution' model instead. However, given how Eq.~\ref{persistence} is set up, and given that positive coefficients in this model indicate link persistence rather than dissolution, we take the liberty to refer to the model as the persistence model.

Since our network is bipartite, we considered links to form between `actors' $i$ (egos) and `events' $j$ (alters). To that end, we employed one endogenous and three exogenous features. Namely, we used the number of edges as the endogenous feature, which controls for network density: $g_1(\mathbf{y},\mathbf{x})=\sum_{ij}y_{ij}=N_e$.  As exogenous features, we included:
\begin{enumerate}
    \item The alters' creative performances (i.e., non-redundant idea counts, $x^{\text{(score)}}$): $g_2(\mathbf{y},\mathbf{x}) = \sum_{ij}y_{ij}x^{\text{(score)}}_j$
    \item Gender-based homophily between the egos and alters: $g_3(\mathbf{y},\mathbf{x}) = \sum_{ij}y_{ij} \mathbb{I}\{x^{\text{(gender)}}_i=x^{\text{(gender)}}_j\}$
    \item Race-based homophily between the egos and alters: $g_4(\mathbf{y},\mathbf{x}) = \sum_{ij}y_{ij} \mathbb{I}\{x^{\text{(race)}}_i=x^{\text{(race)}}_j\}$
\end{enumerate}

where $\mathbb{I}\{\cdot\}$ denotes the indicator function. These four features constitute the network statistic $\mathbf{g}(\mathbf{y},\mathbf{x})$, which is then used in computing the change statistics in the Eqns.~\ref{formation} and ~\ref{persistence}. Note that the fitted coefficients $\beta_f$ and $\beta_p$ are conditional log-odds ratios, so their exponentials can intuitively be interpreted as the factors by which the odds of the formation and persistence of the network ties change respectively. For our implementation, we used the tergm package available within the statnet suite in R, and fitted both formation and persistence models for both of the study conditions.

\subsubsection*{Capturing Inter-ego Semantic Similarity} From each round, we collected pairs of egos who shared (a) $2$ common alters (i.e., exactly the same stimuli), (b) $1$ common alter and (c) no common alter. Within these subgroups, we computed the semantic similarities between every ego-pair's stimulated ideas in turn-2.

To semantically compare the idea-sets of the egos, we first removed stop words and punctuation marks to convert the idea-sets to bag-of-words documents. We represented each document by taking the Word2Vec embeddings of all of the words in the document, and computing the centroid of those embedded vectors. The centroid of a set of vectors is defined as the vector that has the minimum sum of squared distances to each of the other vectors in the set. This centroid is then used as the final document vector representation of the given idea-set~\cite{jurafsky2000speech}. Word2Vec is a popular word-embedding algorithm, which employs skip-gram with negative sampling to train $300$-dimensional embeddings of words~\cite{mikolov2013distributed}.

Given two idea-sets, we computed their document vectors $\mathbf{u}$ and $\mathbf{v}$, and estimated the similarity between the two vectors by taking their cosine similarity,
\begin{equation}
    \text{cosine}(\mathbf{u},\mathbf{v}) = \frac{\mathbf{u \cdot v}}{|\mathbf{u}| |\mathbf{v}|}
\end{equation}

\subsubsection*{Capturing Homogeneity of Ideas within Demographic groups} 
We first considered the sets of ideas that were uniquely submitted by the alters of male and non-male gender identities, but not both. We created vector representations for each of the distinct ideas in the two sets as follows. The same idea can be phrased differently by different people. Therefore, we made use of the manual binnings of ideas described in the Quantifying Creativity subsection, where all the different phrasings of the same idea were collected under a common bin ID. We collected the bin IDs of ideas that were submitted uniquely by males and non-males. Under each bin ID, all the different phrasings of the idea were collected in a bag-of-words document, with all stop-words and punctuation marks removed. Similarly as before, we took the Word2Vec embeddings of the words in this document and computed their centroid to be the final vector representation of the idea.

We then considered pairs of ideas from alters of the same gender, and computed their cosine similarities. However, we only considered idea-pairs from the same round, and if an idea-pair came uniquely from a single person, we ignored that pair. Similarly, we computed the cosine similarities between idea-pairs from alters of different genders, and run statistical tests to confirm homogeneity of ideas within demographic-groups. We ran the race-based idea-homogeneity analysis exactly the same way.

\section*{Data Availability}
Please see https://github.com/ROC-HCI/demography-creativity-networks for the data and code. Due to the copyright protection of the creativity test, we provide processed data of the participants' ideas.

\section*{Acknowledgements}

This work was supported by funding from the National Science Foundation (IIS-1750380), US Army Research Office (W911NF-18-1-0421), and National Institutes of Health (HD-037082).

\section*{Author contributions statement}

R.A.B. designed the study, collected, annotated, and analyzed the data, and authored the manuscript. R.A. contributed to the result interpretation and preparation of the manuscript. G.G. and E.H. oversaw the study design, interpretation of results, and preparation of the manuscript.

\section*{Ethics statement}
The study was reviewed and approved by the Institutional Review Board of the University of Rochester, NY, USA. All participants provided informed consent. All methods were carried out in accordance with relevant guidelines and regulations for involvement of human participants. 

\section*{Additional information}
\textbf{Competing interests} The authors declare no competing interest.

\renewcommand{\thefigure}{S\arabic{figure}}
\renewcommand{\thetable}{S\arabic{table}}
\setcounter{figure}{0}
\setcounter{table}{0}

\clearpage
\date{}


\part*{Supplementary Information}
\textbf{\large Cues to gender and racial identity reduce creativity in diverse social networks}\\

\noindent \author{Raiyan Abdul Baten, Richard Aslin, Gourab Ghoshal, and Ehsan Hoque}\\

\noindent Corresponding Author: Ehsan Hoque,  mehoque@cs.rochester.edu


\maketitle


\clearpage
\subsection*{Capturing Inter-ego Semantic Similarity: Statistical Test Details} 
From each round, we collected pairs of egos who shared (a) $2$ common alters (i.e., exactly the same stimuli), (b) $1$ common alter and (c) no common alter. Within these subgroups, we computed the semantic similarities between every ego-pair's stimulated ideas in turn-2.

We adopted a $3\times2$ factorial design to analyze the data, with $3$ levels in the number of common alters (i.e., the a, b and c subgroups above) and $2$ levels in the study condition factor (i.e., control and treatment). In doing so, we employed the Aligned Rank Transform (ART) procedure, which is a linear mixed model-based non-parametric test. Figure 3 in the main manuscript visualizes the results. We found significant main effects for both of the factors (Number of common alters: $F(2,25964)=135.94$, $P<10^{-15}$; Study condition: $F(1,25964)=369.98$, $P<10^{-15}$). We also found a significant interaction between the two factors ($F(2,25964)=17.81$, $P<10^{-7}$). 

Post-hoc analysis on the ART-fitted model revealed that the semantic similarity between ego-pairs increases as their number of common alters increases ($0$ vs $1$ common alter: $t(25964)=-10.18$, $P<10^{-4}$; $1$ vs $2$ common alter(s): $t(25964)=-9.01$, $P<10^{-4}$). Further pairwise comparisons using $2$-tailed tests showed that this trend holds individually in both of the control and treatment conditions. In the control condition, the inter-ego similarities increased significantly as the number of common alters increased from 0 to 1 and also from 1 to 2 (0 vs. 1 common alter: $t(25964)=-10.61$, $P<10^{-4}$; 1 vs. 2 common alters: $t(25964)=-8.55$, $P<10^{-4}$). The same held for the treatment condition (0 vs. 1 common alter: $t(25964)=-4.71$, $P<10^{-4}$; 1 vs. 2 common alters: $t(25964)=-5.15$, $P<10^{-4}$). These trends intuitively follow the argument that inter-follower similarities can stem from having common stimulation sources. 

Notably, we observed that the inter-ego semantic similarities are significantly higher in the treatment condition compared to their control counterparts, as revealed by post-hoc analysis on the study condition factor in the ART-fitted model ($t(25964)=19.24$, $P<10^{-4}$). Further pairwise comparisons using $2$-tailed tests revealed that this result holds for all of the three common-alter-based subgroups (treatment vs. control; 2 common alters: $t(25964)=5.27$, $P<10^{-4}$; 1 common alter: $t(25964)=13.79$, $P<10^{-4}$; 0 common alter: $t(25964)=17.87$, $P<10^{-4}$). In addition, we observed that with time, the inter-ego semantic similarities in both of the control and treatment conditions show increasing trends (Pearson's correlation, $r$=$0.13$, $P<10^{-4}$ in the treatment condition; $r$=$0.12$, $P<10^{-4}$ in the control condition). All of the $P$-values reported here have been corrected for multiple comparisons using Holm's sequential Bonferroni procedure whenever needed. SI Tables S5-S10 capture these results.

\section*{Supplementary Figures and Tables}
The supplementary figures and tables are listed below.

\clearpage
\begin{figure}
\centering
\includegraphics[width=0.4\linewidth]{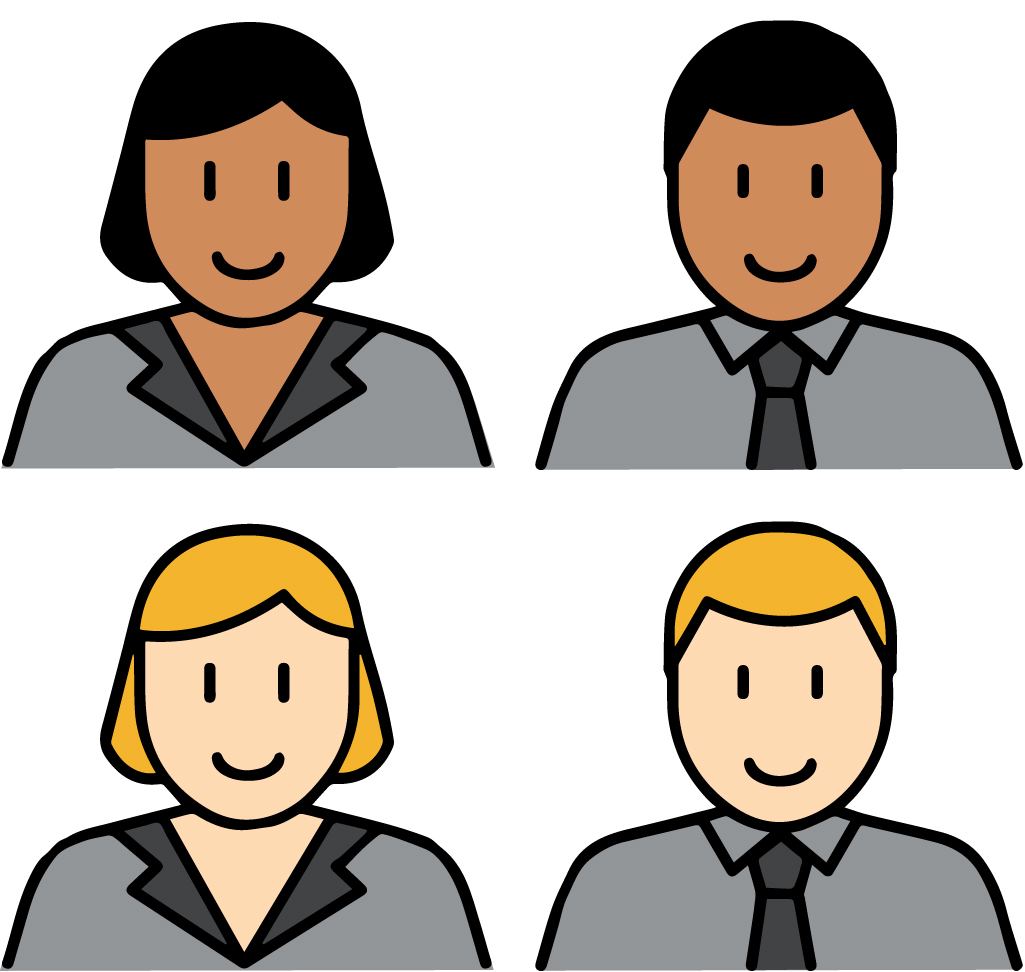}
\caption{Avatars used for depicting demographic information. Top row: Non-White female and Non-White male; Bottom row: White female and White male.}
\end{figure}

\clearpage
\begin{figure}
\centering
\includegraphics[width=1\linewidth]{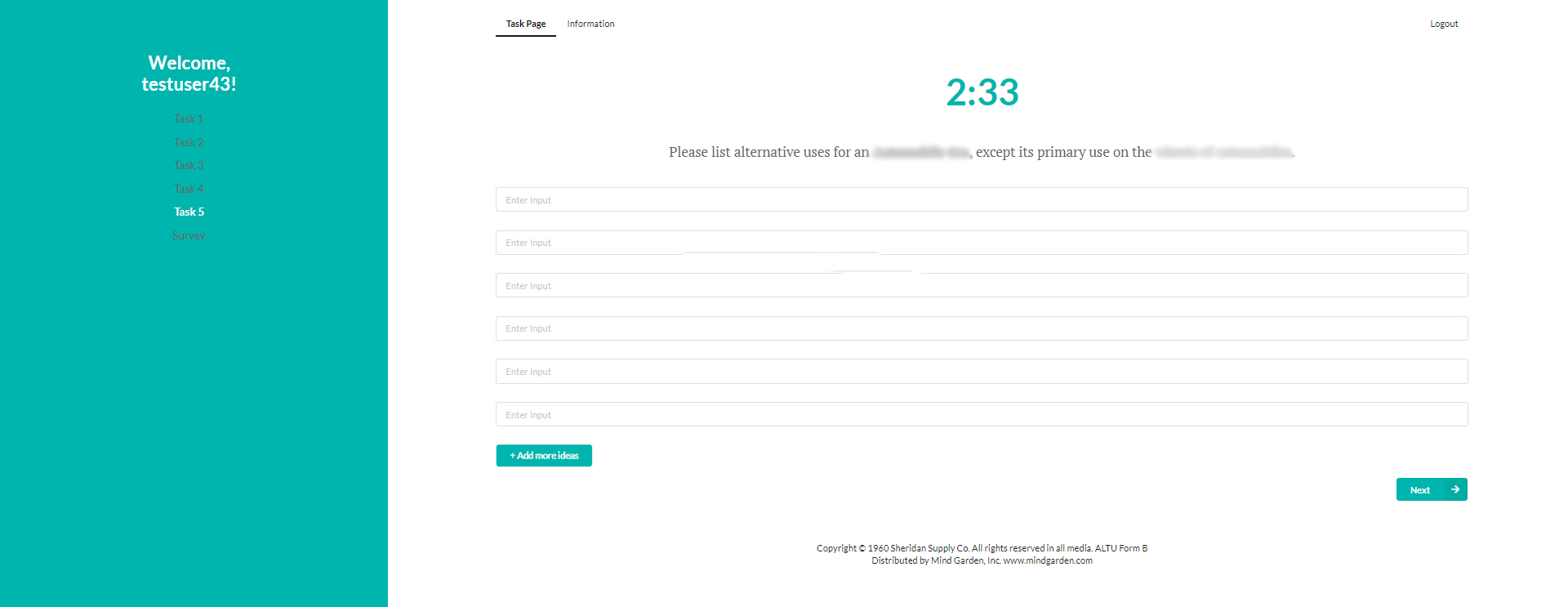}
\caption{Study interface: Initial (turn-1) idea submission interface for egos in both of the study conditions. This interface is used for recording the alters' ideas as well.}
\end{figure}

\clearpage
\begin{figure}
\centering
\includegraphics[width=1\linewidth]{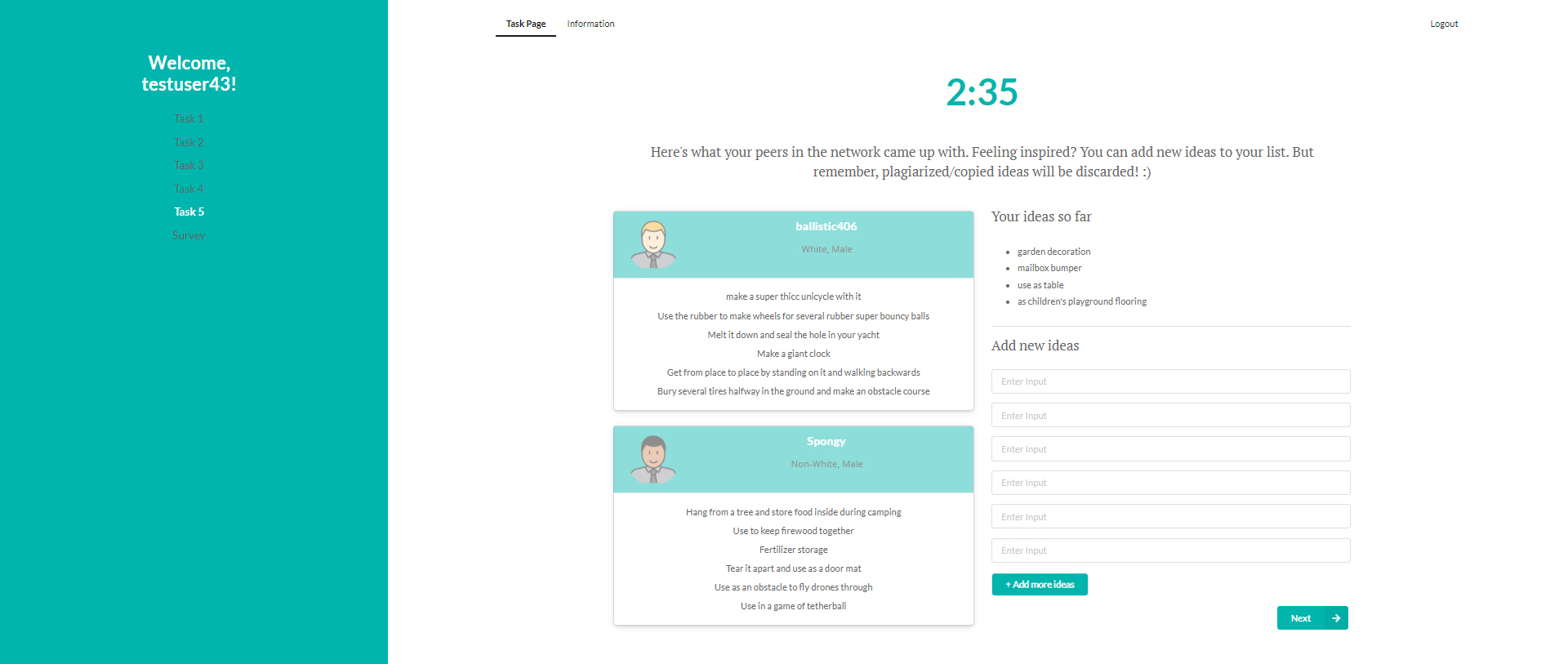}
\caption{Study interface: Turn-2 interface for the egos in the treatment condition. The alters' ideas are shown on the left-side cards. In the control condition, only the usernames and ideas of the alters are shown.}
\end{figure}

\clearpage
\begin{figure}
\centering
\includegraphics[width=1\linewidth]{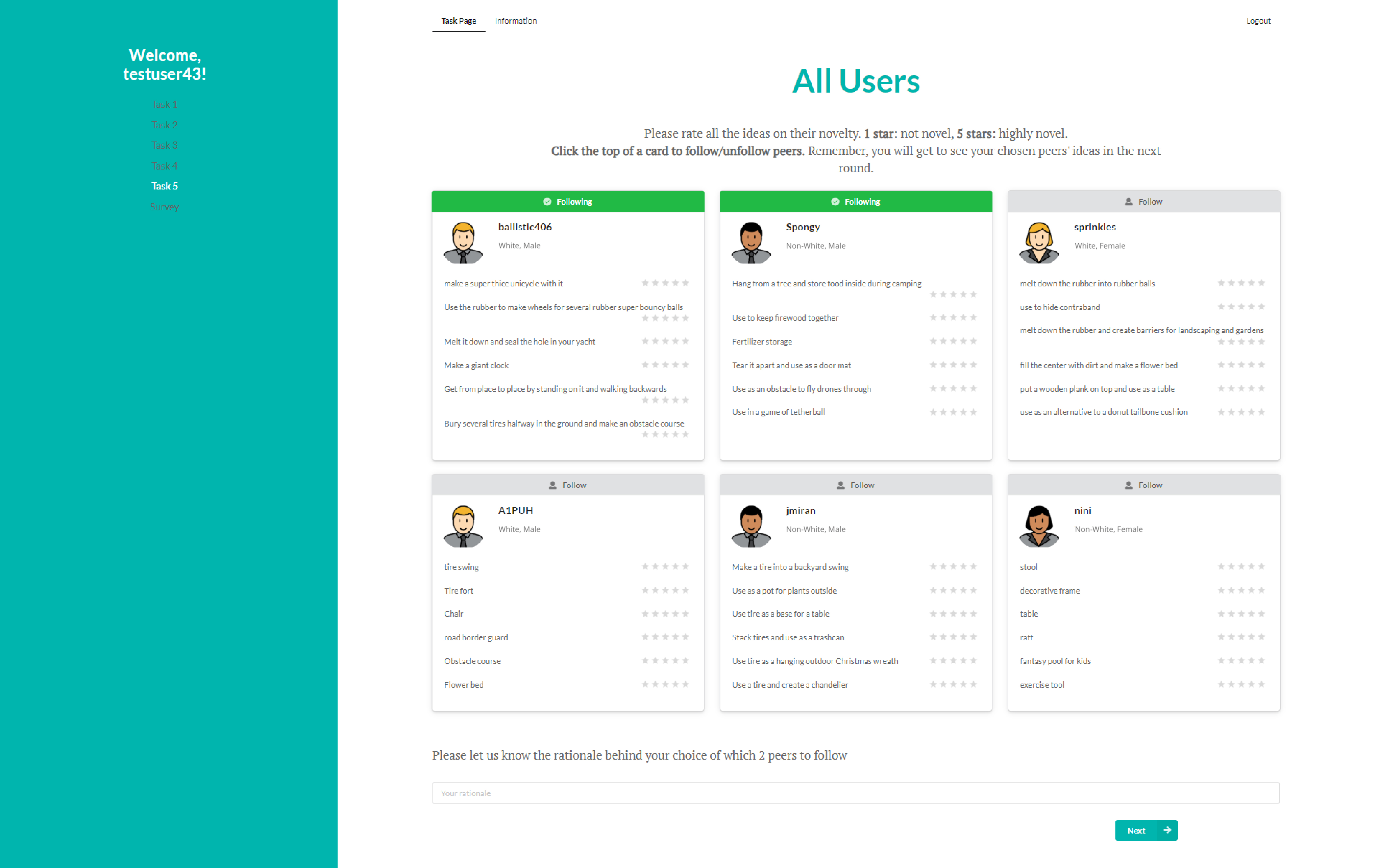}
\caption{Study interface: Rating and rewiring interface for the egos in the treatment condition. In the control condition, only the usernames and ideas of the alters are shown.}
\end{figure}

\clearpage
\begin{table}\centering
\caption{Link formation dynamics in the control condition. Summary results from the Monte Carlo Maximum Likelihood Estimation fit in the STERGM model. ***$P<0.001$}
\begin{tabular}{lrrrrl}
  & $\beta$ & Std. Error & $Z$ value & Pr($>|Z|$) &\\ 
\midrule
Edges                            & $-4.496$ & $0.295$ & $-15.220$ & $<1e-04$ & *** \\ 
Alters' non-redundant idea count & $0.324$  & $0.059$ & $5.505$   & $<1e-04$ & ***\\ 
Gender-based homophily           & $-0.014$ & $0.145$ & $-0.094$  & $0.925$  &\\ 
Race-based homophily             & $0.056$  & $0.144$ & $0.389$   & $0.698$  &\\ 
\bottomrule
\end{tabular}
\label{c_formation}
\end{table}

\clearpage
\begin{table}\centering
\caption{Link persistence dynamics in the control condition. Summary results from the Monte Carlo Maximum Likelihood Estimation fit in the STERGM model. ***$P<0.001$, *$P<0.05$}
\begin{tabular}{lrrrrl}
  & $\beta$ & Std. Error & $Z$ value & Pr($>|Z|$) &\\ 
\midrule
Edges                            & $-0.577$ & $0.290$ & $-1.989$ & $0.0467$& * \\ 
Alters' non-redundant idea count & $0.417$  & $0.060$ & $6.952$   & $<1e-04$& *** \\ 
Gender-based homophily           & $-0.202$ & $0.167$ & $-1.207$  & $0.2275$ & \\ 
Race-based homophily             & $0.179$  & $0.166$ & $1.077$   & $0.2815$ & \\ 
\bottomrule
\end{tabular}
\label{c_persistence}
\end{table}

\clearpage
\begin{table}\centering
\caption{Link formation dynamics in the treatment condition. Summary results from the Monte Carlo Maximum Likelihood Estimation fit in the STERGM model. ***$P<0.001$}
\begin{tabular}{lrrrrl}
  & $\beta$ & Std. Error & $Z$ value & Pr($>|Z|$) &\\ 
\midrule
Edges                            & $-3.740$ & $0.246$ & $-15.230$ & $<1e-04$ & *** \\ 
Alters' non-redundant idea count & $0.197$  & $0.050$ & $3.933$   & $<1e-04$ & ***\\ 
Gender-based homophily           & $-0.117$ & $0.134$ & $-0.879$  & $0.379$  &\\ 
Race-based homophily             & $0.096$  & $0.134$ & $0.714$   & $0.475$  &\\ 
\bottomrule
\end{tabular}
\label{t_formation}
\end{table}

\clearpage
\begin{table}\centering
\caption{Link persistence dynamics in the treatment condition. Summary results from the Monte Carlo Maximum Likelihood Estimation fit in the STERGM model. ***$P<0.001$, **$P<0.01$}
\begin{tabular}{lrrrrl}
  & $\beta$ & Std. Error & $Z$ value & Pr($>|Z|$) &\\ 
\midrule
Edges                            & $-0.751$ & $0.290$ & $-2.588$ & $0.0097$ & ** \\ 
Alters' non-redundant idea count & $0.355$  & $0.058$ & $6.075$   & $<1e-04$ & ***\\ 
Gender-based homophily           & $0.599$  & $0.160$ & $3.743$  & $0.0002$  &***\\ 
Race-based homophily             & $-0.067$ & $0.158$ & $-0.425$   & $0.671$  &\\ 
\bottomrule
\end{tabular}
\label{y_persistence}
\end{table}

\clearpage
\begin{table}\centering
\caption{Omnibus test results for analyzing the inter-ego semantic similarities under various conditions. The cosine similarity between idea-sets of pairs of egos is the response variable. The analysis of variance of Aligned Rank Transformed data is run on a model with two factors: the number of popular alters of the egos ($3$ levels) and the study condition ($2$ levels). The degrees of freedom are specified using the Kenward-Roger method.}
\begin{tabular}{lrrrr}
    & Df & Df.res & $F$ & Pr($>F$) \\ 
\midrule
Number of Popular Alters    & $2$ & $25964$ & $135.944$ & $<2.22e-16$ \\  
 Condition                  & $1$ & $25964$ & $369.983$   & $<2.22e-16$ \\    
 NumPopularAlters:Condition & $2$ & $25964$ & $17.811$   & $1.86e-8$  \\ 
\bottomrule
\end{tabular}
\label{omnibus_}
\end{table}

\clearpage
\begin{table}\centering
\caption{Post-hoc contrast analysis among the three levels in the `number of popular alters' factor from the fitted model reported in Table~\ref{omnibus_}. The degrees of freedom are specified using the Kenward-Roger method. The $P$-values are adjusted using Holm's sequential Bonferroni procedure.}
\begin{tabular}{lrrrr}
 Contrast  & SE & df & $t$ &  $p$ \\ 
\midrule
0 common alter-1 common alter  & $102$ & $25964$ & $-10.176$  & $<0.0001$ \\ 
0 common alter-2 common alters & $141$ & $25964$ & $-15.999$ & $<0.0001$ \\   
1 common alter-2 common alters & $136$ & $25964$ & $-9.014$  & $<0.0001$ \\ 
\bottomrule
\end{tabular}
\label{posthoc-numpopalters}
\end{table}

\clearpage
\begin{table}\centering
\caption{Post-hoc contrast analysis among the two levels in the `study condition' factor from the fitted model reported in Table~\ref{omnibus_}. The degree of freedom is specified using the Kenward-Roger method. The $P$-value is adjusted using Holm's sequential Bonferroni procedure.}
\begin{tabular}{lrrrr}
 Contrast  & SE & df & $t$ &  $p$ \\ 
\midrule
control-treatment  & $104$ & $25964$ & $-19.235$  & $<0.0001$ \\ 
\bottomrule
\end{tabular}
\label{posthoc-condition}
\end{table}

\clearpage
\begin{table}\centering
\caption{Comparisons of cosine similarities among the three levels of `number of popular alters' factor in the control condition. $2$-tailed tests. The $P$-values are adjusted using Holm's sequential Bonferroni procedure.}
\begin{tabular}{lrrrr}
 Contrast  & SE & df & $t$ &  $P$ \\ 
\midrule
0 common alter-1 common alter  & $0.00225$ & $25964$ & $-10.614$  & $<0.0001$ \\ 
0 common alter-2 common alters & $0.00302$ & $25964$ & $-16.067$ & $<0.0001$ \\   
1 common alter-2 common alters & $0.00288$ & $25964$ & $-8.551$  & $<0.0001$ \\ 
\bottomrule
\end{tabular}
\label{posthoc-pairwise_control}
\end{table}

\clearpage
\begin{table}\centering
\caption{Comparisons of cosine similarities among the three levels of `number of popular alters' factor in the treatment condition. $2$-tailed tests. The $P$-values are adjusted using Holm's sequential Bonferroni procedure.}
\begin{tabular}{lrrrr}
 Contrast  & SE & df & $t$ &  $P$ \\ 
\midrule
0 common alter-1 common alter  & $0.00216$ & $25964$ & $-4.709$  & $<0.0001$ \\ 
0 common alter-2 common alters & $0.00312$ & $25964$ & $-8.274$ & $<0.0001$ \\   
1 common alter-2 common alters & $0.00304$ & $25964$ & $-5.152$  & $<0.0001$ \\ 
\bottomrule
\end{tabular}
\label{posthoc-pairwise_treatment}
\end{table}

\clearpage
\begin{table}\centering
\caption{Comparisons of cosine similarities between the two study conditions. $2$-tailed tests. The $P$-values are adjusted using Holm's sequential Bonferroni procedure.}
\begin{tabular}{lrrrr}
 Contrast  & SE & df & $t$ &  $P$ \\ 
\midrule
control-treatment; 2 common alters & $0.00365$ & $25964$ & $-5.271$  & $<0.0001$ \\ 
control-treatment; 1 common alter  & $0.00205$ & $25964$ & $-13.786$ & $<0.0001$ \\   
control-treatment; 0 common alter  & $0.00235$ & $25964$ & $-17.865$  & $<0.0001$ \\ 
\bottomrule
\end{tabular}
\label{posthoc-pairwise_condition}
\end{table}

\end{document}